\documentclass[12pt]{iopart}

\newcommand{\OO}{{\cal O}}

\newcommand{\cd}{\cdot}

\newcommand{\de}{\delta}

\newcommand{\ep}{\epsilon}

\newcommand{\Ga}{\Gamma}

\newcommand{\la}{\lambda}

\newcommand{\ra}{\rightarrow}

\newcommand{\be}{\begin{equation}}
\newcommand{\ee}{\end{equation}}

\newcommand{\bea}{\begin{eqnarray}}
\newcommand{\eea}{\end{eqnarray}}
\newcommand{\bean}{\begin{eqnarray*}}
\newcommand{\eean}{\end{eqnarray*}}

\newcommand{\bk}{{\mathbf k}}

\newcommand{\bB}{{\mathbf B}}
\newcommand{\bA}{{\mathbf A}}
\newcommand{\by}{{\mathbf y}}
\newcommand{\bx}{{\mathbf x}}
\newcommand{\bbe}{{\mathbf e}}

\newcommand{\eg}{{\em e.g. }}

\begin{document}
\title{Primordial Magnetic Fields and Causality}

\author{Ruth Durrer\dag\ and Chiara Caprini\ddag}
\ead{ruth.durrer@physics.unige.ch}
\address{\dag D\'epartement de Physique Th\'eorique, Universit\'e de
  Gen\`eve, 24 quai Ernest Ansermet, CH--1211 Gen\`eve 4, Switzerland}
\ead{caprini@astro.ox.ac.uk}
\address{\ddag Department of Astrophysics, Denys Wilkinson Building, 
  Keble road, Oxford OX1 3RH, UK}


\begin{abstract}
We discuss the implications of causality on a primordial
magnetic field. We show that the residual field on large scales is
much more suppressed than usually assumed, and that a helical
component is even more reduced. Due to this strong suppression, 
even maximal primordial fields generated at the electroweak 
phase transition can just marginally seed the fields in clusters
but they cannot leave any detectable imprint on the cosmic
microwave background.
\end{abstract}

\pacs{98.70.Vc, 98.62.En, 98.80.Cq, 98.80.Hw}

\maketitle

\section{Introduction}

The observed Universe is permeated with large scale coherent magnetic
fields of the order of micro Gauss. It is still under debate whether
these fields have been generated by charge separation processes in the
late universe, or whether primordial seed fields are needed. 
The observational situation is described in Ref.~\cite{kronberg}. More
recent detections of magnetic fields in clusters are discussed in
Ref.~\cite{eilek}.

In this letter we want to
clarify a point which is often missed when investigating cosmic magnetic fields
(for a comprehensive review see~\cite{grasso}): namely that,
because they are divergence-free, magnetic fields are suppressed
stronger than white noise on large scales.

We assume that primordial magnetic fields are generated with a
certain (comoving) coherence scale $L$, by a random
process which is statistically homogeneous and
isotropic. If the field generation occurs
during a non-inflationary phase of the Universe, $L$ must be smaller
than the horizon scale. During inflation, $L$ may diverge and the
arguments presented below do not apply. 
We further assume that the field created by this causal process is
a purely classical magnetic field satisfying Maxwell's equations, or
that it might be considered to be so soon after its generation.
Neglecting possible quantum mechanical fluctuations in the field, 
we can state that its amplitudes and directions must be uncorrelated 
for points which lay farther apart than $L$:
\be  \label{causal}
 \langle B_i(\bx)B_j(\by)\rangle \equiv C_{ij}(\bx-\by) = 0 \:\:\:\:\:\:\:\:\:\:    
\forall ~~ \bx,\by \mbox{ with } |\bx-\by|> L ~. \ee
The first equality comes from the fact that $\bB$ is statistically
homogeneous and isotropic, so that the correlation tensor $C_{ij}$ is 
only a function of $\bx-\by$, and its trace $C=C_{ii}$ depends only
on the distance $r=|\bx-\by|$. For simplicity, we have set the
correlation tensor to zero on separations larger than $L$, but an
exponential decay would  actually be sufficient for all our results.

In Ref.~\cite{hogan} Hogan has argued that the field averaged over
a volume of size $\la^3 >L^3$ behaves like 
\be \label{ho}
B_\la \simeq B_0\left({L\over\la}\right)^{3/2} ~, \ee
where $B_0$ is the amplitude of the field averaged over a volume
given by the correlation scale $L$, and $B_\la$ is the field averaged over 
a volume $V_\la$ of size $\la^3$. Hogan's
argument leading to the above result is very simple: $V_\la$ contains
$N = (\la/L)^3$ uncorrelated volumes. Within each of them the magnetic
field has an average value of $B_0$ pointing in an arbitrary direction. 
The amplitude of the field 
averaged over a volume of size $V_\la$ is therefore reduced by a
factor $\sqrt{N}$, leading to the result~(\ref{ho}).

In this paper we show that this result is not correct and has to be
replaced by
\be \label{we}
B_\la  \simeq B_0\left({L\over\la}\right)^{5/2} ~.
\ee

In what follows we proof Eq.~(\ref{we}). We then apply it to
some relevant cases and show that the difference is important.
We also derive the corresponding scaling behaviour of an helical
magnetic field component. We end with some conclusions.

We shall always use comoving length scales $\la,~L$ and wave
numbers $k$. The scale factor today is normalised to unity,
$a(\eta_0)=1$, $\eta$ denotes conformal time, and we 
assume a spatially flat universe with metric
\[ ds^2 =a^2(\eta)(-d\eta^2+\de_{ij}dx^idx^j) ~.\]
On scales larger than the coherence scale, we consider a field frozen
into the plasma, and simply red-shifting 
with the expansion of the universe like
$\bB(\bx,\eta)=\bB(\bx,\eta_0)a^{-2}(\eta) \equiv
\bB(\bx)a^{-2}(\eta)$. Here $\bB(\bx)$ is the magnetic field scaled to 
its value today, the quantity which we will mainly use from now on. This
behaviour is well justified on 
large scales; on smaller scales however, magnetic energy is converted
into heat due to plasma viscosity. We account for this damping by
introducing a cutoff in the spectrum of the magnetic field
at the smallest scale at which the field is not affected by viscous processes.
A suitable value for the cutoff scale is given 
in Eq. (1) of Ref.~\cite{jedamzik}. If the field has an helicity
component, this model is no longer very appropriate since a process of
inverse cascade may take place \cite{mark}. We will discuss that when
analysing the helicity case.

\section{Causal stochastic magnetic fields}
The fact that magnetic fields are divergence-free implies that the integral
 over an arbitrary closed surface of the normal component of $\bB$ has to
 vanish. This shows that $\bB$ cannot take arbitrary mean
 values in all boxes of size $L$. To see what this implies, let us define the
 Fourier transform,
\[ \hat\bB(\bk) = \int \exp(i\bk\cd\bx)\bB(\bx)d^3x~, \]
so that
\[    \bB(\bx) = {1\over (2\pi)^3}\int \exp(-i\bk\cd\bx)\hat\bB(\bk)d^3k~~.
\]
Since $C_{ij}(\bx)$ is a function with compact support, its Fourier
transform is analytic. If the magnetic field is truly stochastic, 
with no preferred
direction, the only tensors which may enter into the Fourier transform
of its correlation
tensor are combinations of 
$k_n$, $\de_{lm}$ and $\ep_{jlm}$, where $\de_{lm}$ denotes the
Kronecker delta and $\ep_{jlm}$ is the completely antisymmetric tensor in
three dimensions.
The most general Ansatz for the magnetic field correlation tensor in
Fourier space which respects stochastic homogeneity and isotropy is then
\be
 \langle \hat B_l(\bk)\hat B^*_m(\bk')\rangle =
\frac{(2\pi)^3}{2} \delta({\bk}-{\bk'}) 
 [(\de_{lm}-\hat k_l\hat k_m)S(k)   
+ i \epsilon_{lmj} \hat{k}_j A(k)]~.  \label{ansatz}
\ee
With $\hat \bk$ we denote the unit vector in direction of
$\bk$, $\hat\bk=\bk/k$ and $k=|\bk|$. The square bracket in (\ref{ansatz}) is
nothing else than the Fourier transform of $C_{ij}$ and thus has to be
analytic. As we shall see in the next section (see also
Refs.~\cite{cornwall,vachaspati01,pogosian02}), the second term, which changes
sign under the transition $\bk \ra -\bk$ represents a non-vanishing
helicity. We disregard it for this section. 

Causality also implies that $S(k)$
cannot have any structure for values of the wave number smaller than
$L^{-1}$, and hence can be approximated by a simple power law,
\be S(k) = S_0k^n\left(1 +\OO((kL)^2)\right)  \label{Sk} ~.\ee
Analyticity of $(\de_{lm} -\hat k_l\hat k_m)S(k)$ then requires that
\be n\ge 2 \mbox{ is an even integer.} \label{n2}  \ee
Generically, if there are no additional constraints, we 
expect $n=2$. Note that the `usual value' $n=0$ (white noise) is not
allowed because of the non-analytic 
pre-factor $\hat k_l\hat k_m$ which is required to keep the magnetic field
divergence-free, $\bk\cd\hat\bB =0$. In other words, the divergence-free
condition forces a blue spectrum on the magnetic field energy density,
$\langle |\hat\bB(\bk)|^2\rangle\propto k^2$.

Note that this can also be obtained by assuming that the vector potential
$\bA(\bk)$ has a white noise spectrum. With the Ansatz
\be
\langle \bA_i\bA_j^* \rangle =\de_{ij}V(k)+\ep_{ijl}\hat\bk_lW(k)~,
\ee
one has that analyticity requires $W$ to grow at least
like $k$.
Using $\bB = -i\bk\wedge \bA$ one finds $S=k^2V$ and $A=k^2W$.

We want to estimate the average field on a given scale $\la\ge L$. At
this aim, we perform a volume average of the field on a region of size
$\la^3$, following Ref. \cite{hindmarsh}. We convolve $\bB$ with a
Gaussian window function,
\be
\bB_\la(\bx) = {1\over \left(\la\sqrt{2\pi}\right)^3}\int
d^3y\,\bB(\by)\exp\left(-{(\bx-\by)^2\over 2\la^2}\right)~.
\ee

A short computation shows that the magnetic energy density on scale $\la$,
$B_{\la}^2\equiv\langle \bB_{\la}(\bx)^2 \rangle$, is given by 
\be
B^{2}_\la = \frac{1}{(2\pi)^3}\int d^3k\,S(k)\hat f_\la^2(k) =
\frac{S_0}{(2\pi)^2}\frac{1}{\lambda^{n+3}} 
\Gamma\left(\frac{n+3}{2}\right)~, 
\label{B-mean}
\ee
where $\hat f_\la(k) = \exp(-\la^2k^2/2)$ is the Fourier transform
of our window function and $\Ga$ denotes the Gamma function~\cite{Abr}.

For two different scales, $\la_1$ and $\la_2$ we therefore have
\[\left(B_{\la_1}\over B_{\la_2}\right)^2 = \left({\la_2\over \la_1
}\right)^{n+3}~,\] (see also Ref. \cite{hindmarsh}, where however $n=0$
was concluded for causal fields),  
and especially, for the generically expected value $n=2$ 
\be \label{supnorm}
 {B_\la\over B_0} \simeq  \left({L\over\la}\right)^{5/2} ~,
\ee
as claimed in Eq.~(\ref{we}). For $n>2$, the suppression 
with scale is even stronger.

To demonstrate the importance of this additional $L/\la$ factor with
respect to Eq. (\ref{ho}), let us consider
magnetic fields produced during the electroweak phase transition as it
has been proposed by various authors, see \eg \cite{el1,el2,el3,el4} 
(note that the authors of \cite{el4} found a magnetic field spectrum
$\propto k^2$). 
Following these references, at the scale $L_c\sim 10^5$ cm, a
magnetic field with amplitude $B_{ew}(L_c)\sim 10^{-6}$ Gauss is
produced (note that we have scaled the field value to today and we use
conformal length scales normalising the scale factor to unity today,
$a_0=1$). Now $L_c$ is much smaller than 
the horizon scale, $\eta_{ew}\sim 10^{15}$ cm. But in Ref.~\cite{el3}, 
it is argued that  a field is induced also on large
scales, scaling like $L^{-1}$. As we have shown above, this scaling
can only take place up to the horizon scale, and so we have at best
a field of about $B_{ew}(\eta_{ew}) \sim
B_{ew}(L_c)(L_c/\eta_{ew}) \sim 10^{-16}$ Gauss.
As we have argued above, due to causality the field has to decay like
$L^{-5/2}$ on super horizon scales. For $\la \sim 1$ Mpc $\sim 3\times
10^{24}$ cm one can therefore have a field of only about $B_{\la}\sim
10^{-39}$ Gauss, and not $10^{-20}$ Gauss as inferred in
Ref.~\cite{el3} and also in Ref.~\cite{hindmarsh}, where the authors
have set $n=0$ for `frozen-in' magnetic fields.

\section{Helicity}
Let us now investigate limits due to causality on the helicity
component in Eq. (\ref{ansatz}).  
This component can have been produced due to parity violating processes during
the electroweak phase transition, as it has been proposed 
in \cite{cornwall,vachaspati01}. We rewrite the term proportional to
$A(k)$ in Eq.~(\ref{ansatz}) introducing the helicity basis, 
\be 
{\mathbf e}^{\pm }({\bf k}) 
=-\frac{i}{\sqrt{2}}({\mathbf e}_1 \pm  i{\mathbf e}_2)~, 
\label{vector-basis} 
\ee 
where $({\mathbf e}_1,~{\mathbf e}_2,~{\hat \bk})$ 
form a right-handed orthonormal system with ${\bbe}_2 ={\hat \bk}\times
{\bbe}_1$. Setting $\hat\bB(\bk)=B^+\bbe^+ + B^-\bbe^-$ it is
straightforward to see that 
\be
\langle B^{+}({\mathbf k}) B^{+}(-{\mathbf k^\prime)} - 
 B^{-}({\mathbf k}) B^{-}(-{\mathbf k^\prime})\rangle = (2\pi)^3 A(k) 
\delta ({\mathbf k} - {\mathbf k^\prime})~,
\ee
so that $A(k)$ determines the net circular polarisation of the Fourier
mode $\hat\bB(\bk)$.
Again, causality requires that the function
\[ \epsilon_{lmj} \hat{k}_j A(k)\] \
must be analytic and featureless for $k<1/L$, so that
\be A(k) = A_0k^m\left(1 +\OO((kL)^2)\right) ~,\ee
where $m$ has to be a positive odd integer. But there is an additional
constraint coming simply from the Schwarz inequality,
$$ \lim_{\bk'\ra\bk}|\langle (\hat\bk\times \bB(\bk)) \cd \bB(-\bk')\rangle| 
\le \lim_{\bk'\ra\bk}\langle \bB(\bk)\cd \bB(-\bk')\rangle $$ 
implying
\begin{equation}  \label{S>A} 
  |A(k)| \leq S(k) 
\end{equation} 
(note that $S(k) \propto \langle |\hat{\bB}|^2\rangle$, and therefore
 $S(k)\ge 0$). For Eq.~(\ref{S>A}) to be valid for very small values of 
$k$ we must require
\be m \geq n ~. \label{na>ns}\ee
Together with the causality limit from above and (\ref{n2}) this implies
\be \label{indexA}
 m \ge 3 \quad \mbox{ is an odd integer.}  \label{m3}
\ee
Again, generically we expect $m=3$.
Furthermore, applying Eq.~(\ref{S>A}) close to the correlation scale $L$, 
we have 
\be 
 |A_0|  \le S_0L^{(m-n)}~. 
\label{limA} 
\ee 
Vorticity is even more suppressed on large scales by
causality, than a non vortical component of the magnetic field. To
quantify this we define the amplitude of the vortical component on a scale
$\la$ by
\be
 {\mathcal B}^2_\lambda = \frac{\lambda}{(2\pi)^3}\int d^3k\,k\, 
|A(k)|\, \hat{f}_\la^2(k)= 
\frac{|A_0|} {(2\pi)^2}\frac{1}{\lambda^{m+3}}\Gamma 
\left(\frac{m+4}{2}\right)~. 
 \label{ampvort}
\ee 
With the generic value, $m=3$ we therefore have
\be \label{supvort}
 \frac{|{\cal B}_\la|}{|{\cal B}_0|} \simeq  \left({L\over\la}\right)^{3} ~,
\ee
a factor $\sqrt{L/\la}$ more suppression than the non-vortical component, which
for a coherence length of $\eta_{ew}$ translate into an additional
suppression of the order of $10^{-5}$. 
However, for helical magnetic fields an inverse cascade effect takes
place in the early universe, which causes a transfer of power from
smaller to larger scales. This results
in a larger coherence scale than the frozen in one, while the magnetic
spectral index remains unchanged on larger scales \cite{mark,jedamzik}. To
account for this effect, we follow Ref. \cite{vachaspati01}, 
in which the primordial helicity is given by ${\cal H}\sim L_c {\cal
  B}_0^2 =-n_b/\alpha$, where $n_b$ is the baryon density of the
universe today, $n_b(\eta_0) =3\times 10^{-7}$cm$^{-3}$ \cite{sperg},
and $\alpha$ is the fine structure constant. Just as the magnetic field
strength, we have also scaled the helicity, which evolves like $a^{-3}$
to today and is a conserved quantity like the baryon number. 
$L_c$ is the comoving coherence length. According to analytical
studies \cite{Son}, 
the physical coherence scale $aL_c$  evolves with cosmic time roughly like
$t^{2/3}$ (note however that different scaling laws have been found in
numerical simulations, see \cite{mark,jedamzik}). In
Ref. \cite{vachaspati01}, the comoving coherence scale of a maximally 
helical component of the magnetic field is found to be 
$L_c=0.1$ pc. On this coherence scale, the amplitude of the magnetic
field today becomes ${\cal B}_0=\sqrt{n_b/\alpha /L_c}\sim10^{-19}$
Gauss. Taking again a scale $\lambda$ of $1$ Mpc, we get a maximal
helicity amplitude of ${\cal B}_\la={\cal B}_0(L_c/1$Mpc$)^3 \sim
10^{-40}$ Gauss.   

\section{Conclusions}
In this paper we have shown that causally produced magnetic fields
cannot have a white noise spectrum on large scales. They have a blue
spectrum with index $n=2$. A possible helical component of
the field, having a spectral index $m=3$, is even more suppressed on
large scales. These spectral indexes are valid on scales larger than
the coherence scale of the field. The helical component typically has
a larger coherence scale, because of non-linear MHD processing which
leads to an inverse cascade.   

The estimates in the previous sections show that for a magnetic
field causally generated at the electroweak phase transition, 
the amplitudes of the symmetric and helical components at $\la=
1$ Mpc are at most $10^{-39}$ and $10^{-40}$ Gauss: these amplitudes
are too small to seed the magnetic fields observed in
clusters today~\cite{grasso} (see however Ref.~\cite{anne}
which argues that $10^{-30}$ Gauss or even less might suffice).


To answer the question whether we might be able to see some
effects of these fields in the cosmic microwave background
(CMB), we have to estimate the field amplitudes on scales close to about
100 Mpc, which corresponds to an harmonic of about $\ell\sim 400$ (here
we have used the angular diameter distance to the last scattering
surface, $d_A\simeq 13700$ Mpc, from WMAP~\cite{sperg}). 
From the amplitudes for $\la=1$ Mpc given above and using the scaling
behaviour derived in this paper, we obtain
 residual fields of at best $B_{100{\rm Mpc}} \simeq 10^{-42}$ Gauss 
and helicity of ${\cal B}_{100{\rm Mpc}} \sim 10^{-46}$ Gauss on 100 Mpc.
The amplitude of the induced fluctuations in the CMB is typically of
the order of $\de T/T \sim (\rho_B|_{100\rm Mpc})/\rho_\gamma \ll
10^{-5}$.

We therefore conclude, that a magnetic field which has evolved on large
scales simply via flux conservation from its creation at the
electroweak phase transition until today is not sufficient to
have seeded the large scale magnetic fields observed in 
clusters, even if a dynamo mechanism could amplify it during the
process of structure formation. The same conclusion can be made
for an helicity component of the magnetic field, if accounting for MHD
processing in the simple way as explained in the previous paragraph. 
Such a field also does not lead to
observable traces in the anisotropies or the polarisation of the CMB. 
This latter conclusion has also been drawn in previous
works~\cite{pedro,hewaves}.  

Possible ways out are either that the magnetic fields observed in clusters are 
due to very small scale seed fields, coherent on scales of
the order of a parsec or less. Another possibility is that the seed
fields have been  generated by a `non-causal' mechanism, \eg during an
inflationary phase, see \cite{tw,gasp}. But also in this latter case,
a very red spectrum $n<-2$ is needed for the magnetic fields to play
at the same 
time the role of seeds for large scale magnetic fields and to lead to
visible imprints on the CMB. Such red spectra are actually also required by the
limits from small scale gravitational waves which are induced by
magnetic fields~\cite{cr}. 

 Our results strongly disfavour large scale seeds
induced from small scale coherent magnetic fields which might be produced in
the early Universe.
\vspace{0.5cm}

\ack
We acknowledge discussions with Karsten Jedamzik, Pedro Ferreira and
Tina Kahniashvili. C.C. thanks Geneva University for hospitality. This
work is supported by the Swiss National Science Foundation.


\section*{References}
\begin{thebibliography}{99}
\bibitem{kronberg}P. Kronberg, Rep. Prog. Phys. {\bf 57}, 57 (1994).
\bibitem{eilek}J. Eilek and F. Owen, Astrophys. J. {\bf 567}, 202 (2002).
\bibitem{grasso}D. Grasso and H.R.~Rubinstein, Phys.\ Rept.\ {\bf
  348}, 163 (2001).  
\bibitem{hogan}C. Hogan, Phys. Rev. Lett. {\bf 51}, 1488 (1983).
\bibitem{jedamzik}R. Banerjee and K. Jedamzik, preprint {\tt astro-ph/0306211} 
\bibitem{mark} M. Christensson, M. Hindmarsh and A. Brandenburg,
  Phys. Rev. {\bf E64}  056405 (2001).
\bibitem{cornwall}J. M. Cornwall, Phys. Rev. {\bf D56}, 6146 (1997) 
\bibitem{vachaspati01} T.~Vachaspati, Phys. Rev. Lett. {\bf 87} 
  251302 (2001). 
\bibitem{pogosian02}  L.~Pogosian, T.~Vachaspati, S.~Winitzki, 
Phys.\ Rev.\ D {\bf 65}, 3264 (2002). 
\bibitem{hindmarsh} M. Hindmarsh and A. Everett, Phys. Rev. {\bf D58},
  103505 (1998). 
\bibitem{Abr} M.~Abramowitz and I.~Stegun, 
{\it Handbook of Mathematical Functions} (Dover, New York, 1972). 
\bibitem{el1}T. Vachaspati, Phys. Lett. {\bf B265}, 258 (1991).
\bibitem{el2}M. Joyce and M. Shaposhnikov, Phys. Rev. Lett. {\bf 79},
  1193 (1997).
\bibitem{el3}J.T. Ahonen and K. Enqvist, Phys. Rev. {\bf D57}, 664 (1998).
\bibitem{el4}D. Boyanovsky, H. J. de Vega and M. Simionato, preprint {\tt astro-ph/0305131} 
\bibitem{sperg}D. Spergel et al., preprint {\tt astro-ph/0302209}.
\bibitem{Son}D.T. Son, Phys. Rev. {\bf 59}, 063008 (1999)  
\bibitem{anne}A. Davis, M. Lilley and O. T\"ornkvist, Phys. Rev. {\bf
  D60} 021301 (1999).
\bibitem{pedro} R.~Durrer, P.G.~Ferreira, and T.~Kahniashvili, 
Phys.\ Rev.\ D {\bf 61}, 043001 (2000). 
\bibitem{hewaves}C. Caprini, T. Kahniashvili and R. Durrer,
  Phys. Rev. D  submitted (2003).
\bibitem{tw}E. Turner and L. Widrow Phys. Rev. {\bf D37} 2743 (1988).
\bibitem{gasp}M. Gasperini, M. Giovannini and G. Veneziano,
  Phys. Rev. Lett. {\bf 75} 3796 (1995).
\bibitem{cr}C. Caprini and R. Durrer, Phys. Rev. {\bf D65} 023517 (2002).
\end{thebibliography}
\end{document}